# A Vision of DevOps Requirements Change Management Standardization

Muhammad Azeem Akbar[1,*], Arif Ali Khan[2], Sajjad Mahmood[3, 4], Saima Rafi[5]
[1]Department of Software Engineering, LUT University, Lappeenranta, Finland.
[2]M3S Empirical Software Engineering Research Unit, University of Oulu, 90570 Oulu, Finland.
[3]Information and Computer Science Department, King Fahd University of Petroleum and Minerals, Dhahran, Saudi Arabia.
[4]Interdisciplinary Research Center for Intelligent Secure Systems, King Fahd University of Petroleum and Minerals, Dhahran, Saudi Arabia.
[5]University of Murcia, Department of Informatics and Systems, Murcia, Spain.
azeem.akbar@ymail.com, arif.khan@oulu.fi, smahmood@kfupm.edu.sa, saeem112@gmail.com
*Muhammad Azeem Akbar

*Abstract*— DevOps (development and operations) aims to shorten the software development process and provide continuous delivery with high software quality. To get the potential gains of DevOps, the software development industry considering global software development (GSD) environment to hire skilled human resources and round-the-clock working hours. However, due to the lack of frequent communication and coordination in GSD, the planning and managing of the requirements change process becomes a challenging task. As in DevOps, requirements are not only shaped by development feedback but also by the operations team. This means requirements affect development, development affects operations and operations affect requirements. However, DevOps in GSD still faces many challenges in terms of requirement management. The purpose of this research project is to develop a DevOps requirement change management and implementation maturity model (DevOps-RCMIMM) that could assist the GSD organizations in modifying and improving their requirement management process in the DevOps process. The development of DevOps-RCMIMM will be based on the existing DevOps and RCM literature, industrial empirical study, and understanding of factors that could impact the implementation of the DevOps requirement change management process in the domain of GSD. This vision study presents the initial results of a systematic literature review that will contribute to the development of maturity levels of the proposed DevOps-RCMIMM.

## I. INTRODUCTION

DevOps refers to a multidisciplinary software development culture, where the knowledge workers of the development and operations team work in a close collaborative fashion [1]. DevOps help in building, and managing a scalable and secure platform by creating automated pipelines (that assist in building software and infrastructure more efficiently)[2]. DevOps is all about meeting the demands of customers in a secure and continuous way[1]. Due to the instant feedback in DevOps, the requirements change frequently.

The change management in the DevOps process is a complicated activity of the requirements engineering phase as it is hard to assess the scope of complete software due to the evolving needs of the client [3]. Nurmuliani et al.[3] underlined that requirements change management (RCM) is "the tendency of requirements to change over time reacting to the evolving needs of customers, stakeholders, firms, and the work environment." Moreover, they underlined that the continuous business and technology change forces the client to move forward to propose a change request. Therefore, a guidelines-based model is required for requirement change management in DevOps.

Various models and frameworks for RCM have been introduced to support software development firms in controlling the changes in software requirements. Ince [4] developed an RCM model and considered the basic aspect of change management. However, Ince's RCM model lacks with verification mechanism for demanded changes. Therefore, it is challenging to evaluate whether the changes made in the system are working accordingly or not. Similarly, Keshta et al. [5] develop an RCM model to manage the change management activities in small and medium software development organizations. The model covers the change management aspect of the traditional software development paradigm and lacking to deal with the DevOps environment. Akbar et al.[6] also, develop a change management maturity model and readiness model [7] that provides a roadmap to address the RCM in the GSD context, but did not cover the continuous software engineering paradigm.

These frameworks have effectively addressed the activities of the RCM process but ignore the continuous software engineering characteristics i.e., delivery, development, and deployment. Therefore, this research gap motivated us to develop a model that will assist DevOps in GSD organizations in modifying and improving their requirement management and development process in DevOps firms. This study will give an important understanding of the RCM in DevOps process challenges and success factors in the domain of GSD.

## II. RESEARCH QUESTIONS AND OBJECTIVES

As per existing literature, the is a dire need of a roadmap to address the RCM challenges in DevOps and GSD. The plan of this project is to understand the RCM process in the context of DevOps in GSD with the research community and software industry experts; considering the opinions of software engineering research and practitioner, to develop a maturity model that will give assistance to the evaluation and improvements of change management activities in DevOps and GSD context. Thus, to come up with this research project, the following research questions (RQs) are proposed:

**RQ1:** What success factors and barriers of globally distributed DevOps requirements change management program are reported in literature?
**RQ2:** What success factors and barriers of globally distributed DevOps requirements change management program are investigated in the industrial empirical study?
**RQ3:** Are there any differences exists between the findings of literature and industrial empirical study?
**RQ4:** What are the best practices to fix the explored success factors and barriers?
**RQ5:** How can a robust DevOps requirement change management and implementation maturity model (DevOps-RCMIMM) be developed?

### III. RESEARCH METHODOLOGY

This research project is empirical in type and to understand the concept of software engineering research community towards RCM-DevOps in the GSD context, we will apply the systematic literature review (SLR). As the SLR is a systematic way to explore the potential literature in-line with the proposed RQs, and it is a widely adopted research approach in software engineering domain [8]. In this project, the key aim of considering SLR is to explore the factors that could negatively and positively influence the RCM-DevOps in GSD context, reported by research community [9]. To understand and get the industrial experts insight concerning to the project objective, we will conduct questionnaires survey and interviews. At final stage, we will consider case study method to evaluate the implacability and suitability of proposed model (DevOps-RCMIMM) in software industry, as the case study is considered to be the most influential evaluation tool used by other researchers in software engineering [10, 11] and it provides real world information [12].

Therefore, for this preliminary study, we used the following SLR protocols to identify the success factors and challenges reported by research community concerning to RCM in DevOps and GSD.

*A. Phase 1: Planning the review*

*Research questions (RQs):* The purpose of this paper is to identify the factors that could have positive and negative impact on RCM-DevOps process in GSD environment. This study is conducted to address only RQ1, as presented in section II.

*Data sources:* The appropriate data sources were selected based on our understanding and recommendations provided by Chen et al. [13], Khan et al. [14] and Inayat et al. [15]. The selected digital repositories are listed in Table 2:

*Search strings:* To extract the most potentially relevant literature from the selected repositories, the key terms were used. Using the keywords and their alternative derived from the existing DevOps, RCM and GSD literature, search string was formulated [5, 10, 14, 16]. We further used "OR", "AND" operators to concatenate and to formulate the complete search strings, for instance:

("success factors") AND ("barriers") AND ("requirements change managements") AND ("DevOps") AND ("global software development")

*Inclusion criteria:* To develop the literature inclusion criteria, we used the protocols given in [14, 15, 17].
- The selected piece of literature should be a conference, journal article or book chapter.
- Literature to elaborate RCM practices in DevOps and GSD.
- Articles that reported the negative and positive influencing factors of DevOps RCM process in GSD context.
- The studies based on empirical evaluation have more significance.

*Exclusion criteria:* We used the criteria developed by Khan et al.[14] to exclude the studies which were not relevant to our research objectives.
- Studies that do not describe the details of DevOps and RCM practices in GSD.
- Study other than English.
- The pre-mature studies were also excluded.

*Study Quality Evaluation (QE):* During final literature selection process, the QE process was performed with the intent to check the significance of selected literature to address the proposed research question of our study. To determine the quality of the selected literature, we developed a QE checklist (Table 1). The QE criteria were developed using the guidelines given in [14, 18]. The checklist consists of five questions (QE1-QE5). For each question, the evaluation was done as follows:
- "If an article fully covers the objective of proposed study, then assigned score was 1".
- "If a paper partially full the RQs, then assigned- score 0.5".
- "If a criterion was not fulfilled, then then assigned- score 0".

Table 1: Study Quality Evaluation Criteria

| QE | Checklist Questions |
| --- | --- |
| QE1 | Are the used research methods address the proposed RQs? |
| QE2 | Do the selected study discuss the RCM in DevOps and GSD? |
| QE3 | Does the change management process explicitly address? |
| QE4 | Do the study cover the GSD concept for RCM in DevOps? |
| QE5 | Does the results of literature enough to address the proposed RQs? |

*B. Phase 2: Conducting the review*

*Primary study selection:* The selected studies were refined by applying the tollgate approach developed by Afzal et al. [18], which consists of total 5 phases (Table 2):

Initially, 2366 paper were collected by executing the search terms on selected data sources given in Phase-1 respectively. We have finally selected 107 primary studies using tollgate approach [18], which shows 4.52% of the total articles (Table 2). Finally, the selected primary studies were evaluated using the QE criteria (discussed in Phase 1). The list of the primary studies is given at https://tinyurl.com/yc6pxjap and each primary study is labelled as [SP] to consider it as the primary study.

Table 2: Tollgate Approach

| Digital source | P-1 | P-2 | P-3 | P-4 | P-5 | % (N=107) | Search Date |
|---|---|---|---|---|---|---|---|
| ACM Digital Library | 263 | 101 | 60 | 29 | 11 | 10 | 05-04-2022 |
| IEEE Xplore | 624 | 511 | 196 | 90 | 29 | 27 | 13-04-2022 |
| Wiley Inter Science | 273 | 165 | 84 | 31 | 10 | 8 | 16-04-2022 |
| Springer Link | 257 | 129 | 71 | 35 | 13 | 12 | 20-04-2022 |
| Science Direct | 345 | 308 | 102 | 32 | 11 | 10 | 03-05-2022 |
| IET Digital Library | 81 | 78 | 23 | 16 | 7 | 7 | 11-05-2022 |
| Google Scholar | 523 | 484 | 182 | 117 | 26 | 25 | 30-09-2022 |
| Total | 2366 | 1776 | 718 | 350 | 107 | 100 | |

*Data extraction:* The data extraction team consist of all the authors of this study. Firstly, the author no.1 and 4 extract the success factors and barriers from reported in the selected literature. For further verification, author no. 2 and 3 continuously participated to check the baseness and consistency of data extraction process. In order to check the consistency and correctness of data extraction process, we further execute- inter-personal by using reliability test, and for these two external experts are invited. The external experts randomly select 10 papers and perform data extraction with same sequence to authors. Considering the extracted data from 10 articles, we determine the non-parametric Kendalls coefficient of concordance (W) with the aim to check the agreement with both set of extracted data i.e., extracted by study authors and externals [19, 20]. The determined W=0.89 (p=0.002), shows the strong and positive agreement in both data set. Thus, based on the determined W, we assumed that the data extraction process is consistent and unbiased.

*C. Phase 3: Reporting the review*

*Quality check of selected set of literature:* The selected literature was evaluated using the QE criteria given in Table 1. The evaluated QE score along with each selected piece of literature is given at https://tinyurl.com/yc6pxjap. The QE score shows 78% of the selected literature scored ≥ 60%, which shows the potential of selected literature to answer the RQs of our study [14, 17, 21]

*List of identified success factors and barriers:* The list of identified DevOps RCM success factors and barriers are given in Table 3 and Table 4.

## IV. STRUCTURE of DevOps-RCMIMM

The conceptual architecture or proposed model is presented in Figure 1. The basic structure is based on the other model of software engineering domains like CMMI, IMM and SOPM [19, 22, 23] and influencing factors of DevOps-RCM in GSD context. The proposed structure comprises of three core modules including (critical success factors (CSFs) critical barriers (CBs) and assessment component. The association among all the components and their interrelationship is presented in Figure 1. To assess the capability of an organizations concerning to RCM-DevOps in GSD context, assessment component will be used. The factors components educate the practitioners concerning to the critical positive and negative influencing areas of RCM-DevOps in GSD context. Moreover, the assessment component will assist to choose the best practices in order to improve the DevOps-RCM process in GSD context.

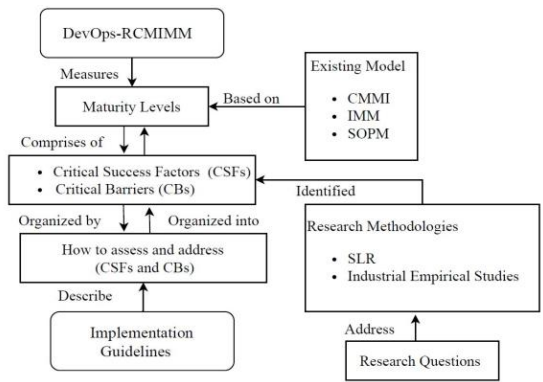

Figure 1. Structure of DevOps-RCMIMM

## V. PRELIMINARY SLR RESULTS

*D. Findings of SLR*

In this current study, we used the SLR approach to explore the success factors and barriers of RCM-DevOps in GSD context. Thus, by using the SLR protocols, we identified 22 success factors that could positively influence and 20 barriers that could negatively impact the RCM-DevOps in GSD. The objective of identifying success factors and barriers is to come-up with the DevOps-RCMIMM factors component as discussed in section 4. The lists of success factors and barriers are collected by reviewing selected 107 primary studies, which are given at: https://tinyurl.com/yc6pxjap. The identified success factors and barriers are presented in Table 3 and 4, respectively.

Table 3 Success factors identified using SLR approach

| S. No | Investigated success factors | S. No | Investigated success factors |
|---|---|---|---|
| SF1 | DevOps change management engineering | SF12 | Frequent overseas site's response |
| SF2 | Sharing information across distributed DevOps sites | SF13 | Change management automation in DevOps |
| SF3 | Requirement's elicitation and identification with DevOps teams | SF14 | Effective RCM leadership in DevOps |
| SF4 | Effort and cost estimation for change | SF15 | DevOps top management commitment |
| SF5 | Clear change management strategy in DevOps | SF16 | DevOps RCM team motivation |
| SF6 | Change management process awareness among DevOps team | SF17 | Accountability of RCM activities |
| SF7 | Governess and control of RCM activities | SF18 | Determining the impact of demanded changes on DevOps |
| SF8 | Vision and goal for change | SF19 | Proper RCM task coupling in DevOps |
| SF9 | Advance & uniform RCM infrastructure at DevOps GSD sites | SF20 | RCM process measurement across DevOps GSD sites |
| SF10 | Financial stability in DevOps | SF21 | Understanding of requested changes in DevOps |
| SF11 | Change identification and validation in DevOps | SF22 | Frequent communication and dialog in DevOps |

Table 4. Barriers identified using SLR approach

| S. No | Investigated challenge | S. No | Investigated challenge |
|---|---|---|---|
| B1 | Requirements tracking and control issues in DevOps | B11 | Lack of resources management across distributed DevOps sites |
| B2 | Inexperienced RCM staff in DevOps | B12 | Unclear scope of requested changes in DevOps |
| B3 | Unavailability of geographically distributed CCB (change control board) | B13 | Lack of change management planning in DevOps activities |
| B4 | RCM risk identification and management issues in DevOps | B14 | Lack of change impact analysis in distributed DevOps sites |
| B5 | Unavailability of RCM standards in DevOps process | B15 | Different rules and policies of distributed DevOps sites |
| B6 | Time and budget constraints for RCM process in overall DevOps | B16 | Lack of work synchronization between different DevOps sites |
| B7 | Poor DevOps organizational infrastructure | B17 | Delay in response from offshore DevOps teams |
| B8 | Lack of requirements analysis techniques in DevOps | B18 | Lack of 3C's (communication, coordination and control) in DevOps practitioners |
| B9 | Lack of DevOps leadership support | B19 | Lack of RCM process training |
| B10 | Lack of domain Knowledge | B20 | Requirements remain too abstract in DevOps |

*E. Critical Factors*

Niazi et al. [21] underlined that the top organizations management should consider urgently and seriously, the factors declared as most critical, with the intent to gain the expected business objectives. If less consideration given to the critical factors could have serious consciences on business gains [14, 19]. Considering other software engineering research studies, we used the following criteria to identify the critical success factor and barriers of RCM-DevOps in GSD context.

The factors have frequency ≥ 50% from the selected primary studies, is ranked as critical. The frequency analysis of the reported factors is provided at https://tinyurl.com/yc7a39h7. Using this criteria, the success factors: CSF1, CSF6, CSF7, CSF9, CSF16, and CSF17 are declared as the critical factors. Similarly, the following barriers are reported as the critical barriers: CB1, CB3, CB5, CB6, CB9, CB14, CB18 and CB19.

*F. Categorical Classification of the Investigated Success Factors and Barriers*

Ramasubbu [24] develop a framework considering success factors of software process improvement, to indicate the domain based critical success factors. Following the Ramasubbu framework, other software engineering studies also taxonomies the success factors of software process improvements [9, 25]. Using Ramasubbu [24], we develop a framework by mapping the identified success factors, barriers based on experience of the research team. This informal classification will further be validated in future by getting insight of industry practitioners.

According to the mapping results, majority of the identified success factors and barriers are belonging to "project administration" category (Figure 2). This shows that "project administration" is the most critical area of RCM-DevOps in GSD context, and that need more intention from industry experts. The developed framework given a base knowledge for software engineering research and practitioners community to develop new guidelines and framework to successfully execute the RCM-DevOps in GSD context. The framework given Figure 2, will help them to consider more important areas on priority, considering the mapped success factors and barriers.

## VI. MATURITY LEVEL COMPONENTS of DevOps-RCMIMM

We have followed the structure of CMMI [22] to develop the maturity levels of DevOps-RCMIMM. The levels of CMMI comprises of 25 process areas (PAs) that have consisted of various real-world practices. The PAs are mapped in five maturity levels [22]. Thus, to develop the DevOps-RCMIMM levels, we will map the barriers and success factors instead of PAs. In different other maturity models (i.e., IMM, SOPM), the concept of critical success factors (CSFs) and critical barriers (CBs) was adopted. They have reported the significance of CSFs and CBs. We have followed the same concepts and preliminary develop the maturity levels of the proposed model (DevOps-RCMIMM) based on the CSFs and CBs investigated during the SLR study (Figures 3) that are given at https://tinyurl.com/y98xp4ys.

The maturity levels will be further modified based on the industrial empirical study that will be conducted in the future. A brief description of proposed DevOps-RCMIMM maturity levels (Figure 3) are as follows:

*Level-1 (Initial):* The level-1 of CMMI has been directly considered as the initial level of DevOps-RCMIMM. In this level, the deployment process of DevOps RCM is chaotic, and few processes are defined. This level has no CSF and CB.

*Level-2 (Managed):* This level is directly derived from CMMI and at this level all activates of DevOps RCM process are properly managed. This level includes one CSF and two CBs.

*Level-3 (Change management control):* This level of DevOps-RCMIMM is derived from a most significant reported barrier in the selected primary studies (i.e., lack of geographically distributed change control board). The demanded requirements changes could affect the development activates carried at overseas sites. Therefore, change management control is significant to manage the demanded requirements changes effectively and efficiently at GSD sites in DevOps process. Therefore, this level includes two CSFs and three CBs.

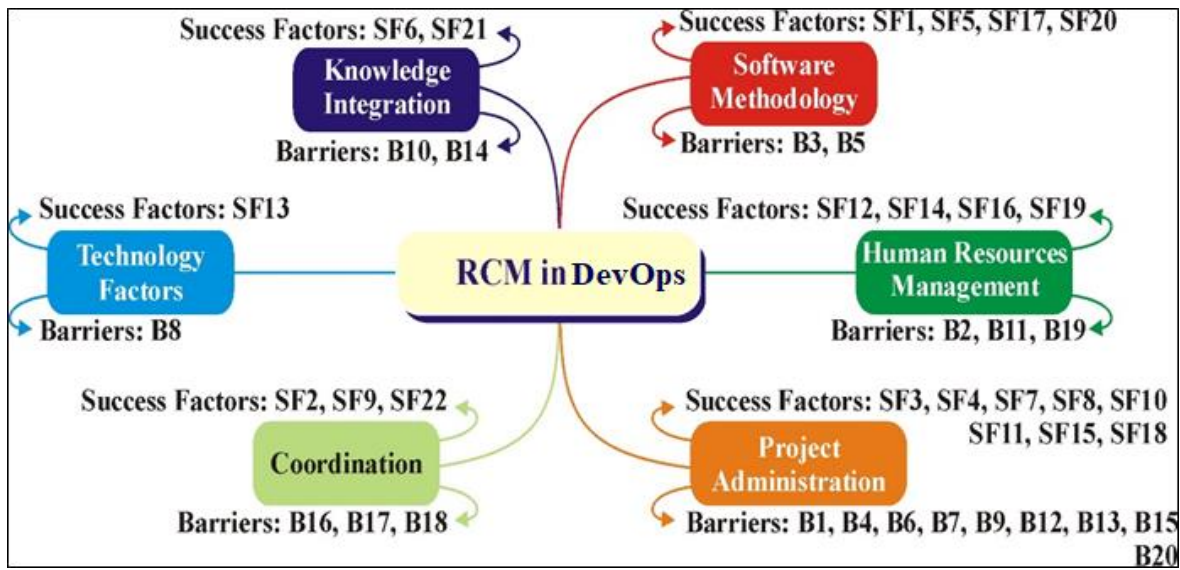

Figure 2. Categorization of identified success factor and barriers

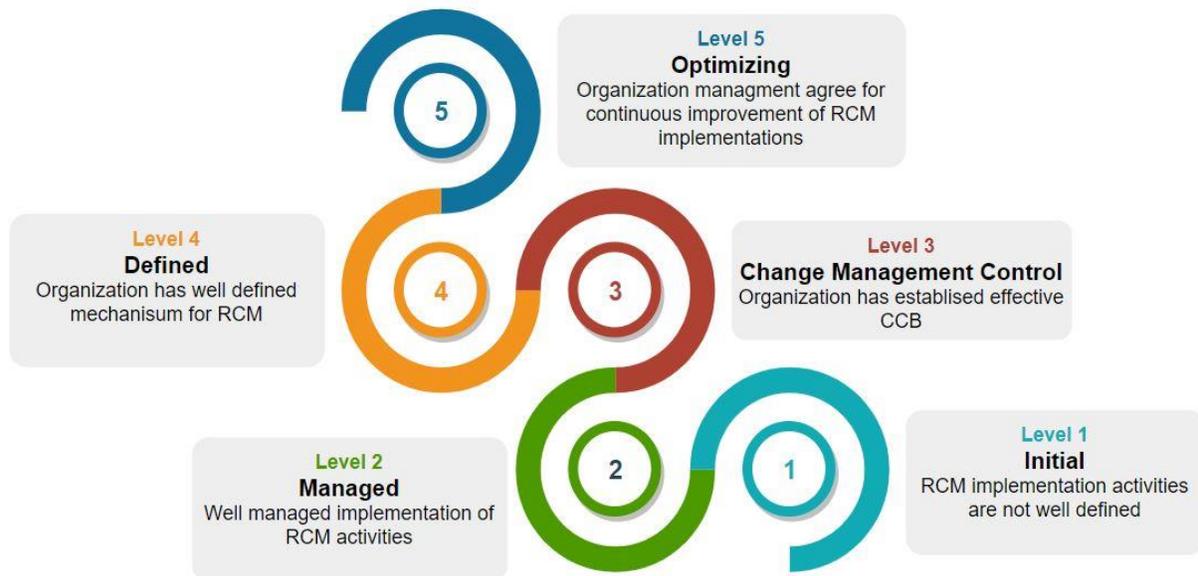

Figure 3. DevOps-RCMIMM maturity levels

*Level-4 (Defined):* Level-4 has been adopted from CMMI as in this level all the processes of DevOps RCM implementation are reported and integrated into a standard implementation process for the organization. This level has two CSFs and one barrier.

*Level-5 (Optimizing):* Optimizing is the final maturity level of DevOps-RCMIMM, and it is directly adopted from CMMI. This level leads towards continuous RCM in DevOps in GSD context.

DevOps-RCMIMM does not adopt CMMI level-4 (Quantitatively Managed). The focus of level-4 (Quantitatively Managed) is on assessing the quantitative characteristics of the DevOps RCM process. In this research work, we didn't identify any success factor or barrier that has direct association with the level-4 CMMI.

## VII. RESEARCH CONTRIBUTION AND FUTURE WORK

The ultimate aim of this project is to develop a maturity model for the successful execution of DevOps based RCM process in GSD context. The proposed model will assist the practitioners in evaluating and improving the maturity level of DevOps organization while performing the RCM activities. This is the first study towards the exploration of RCM influencing factors in the context of DevOps and GSD. The current study only explores the success factors and challenges of DevOps RCM in GSD context using SLR which will contribute to the development of maturity level of proposed model i.e., DevOps-RCMIMM.

The future of this research project is to understand the practitioners' opinions concerning to DevOps-RCMIMM in GSD context. For this, will use different empirical investigation techniques to get the practitioners insight using survey instrument, case study, and interview as well. The final stage of this project is to come-up with maturity levels and design and complete DevOps-RCMIMM.